 \documentclass[final,3p,times,12pt]{elsarticle}

\usepackage{amsmath}
\usepackage{amssymb}
\usepackage{bm}
\usepackage[english]{babel}
\usepackage[dvips]{color}
\usepackage{graphicx}

\journal{Nuclear Physics A}

\begin{document}

\begin{frontmatter}

%% Title, authors and addresses

%% use the tnoteref command within \title for footnotes;
%% use the tnotetext command for theassociated footnote;
%% use the fnref command within \author or \address for footnotes;
%% use the fntext command for theassociated footnote;
%% use the corref command within \author for corresponding author footnotes;
%% use the cortext command for theassociated footnote;
%% use the ead command for the email address,
%% and the form \ead[url] for the home page:
%% \title{Title\tnoteref{label1}}
%% \tnotetext[label1]{}
%% \author{Name\corref{cor1}\fnref{label2}}
%% \ead{email address}
%% \ead[url]{home page}
%% \fntext[label2]{}
%% \cortext[cor1]{}
%% \address{Address\fnref{label3}}
%% \fntext[label3]{}

\title{Quarks Production in the Quark-Gluon Plasma
Created in Relativistic Heavy Ion Collisions}

\author[label1]{M. Ruggieri\corref{cor1}}
\ead{marco.ruggieri@lns.infn.it}
\author[label1,label2]{S. Plumari}
\author[label1,label2]{F. Scardina}
\author[label1,label2]{V. Greco}

\address[label1]{Department of Physics and Astronomy, University of Catania, Via S. Sofia 64, I-95125 Catania}
\address[label2]{INFN-Laboratori Nazionali del Sud, Via S. Sofia 62, I-95123 Catania, Italy}

%\address{}
\cortext[cor1]{Corresponding author.}

\begin{abstract}
In this article we report on our results about quark production and chemical equilibration of quark-gluon plasma.
Our initial condition corresponds to a classic Yang-Mills spectrum, in which only gluon degrees of freedom
are considered; the initial condition is then evolved to a quark-gluon plasma by means of relativistic
transport theory with inelastic processes which permit the conversion of gluons to $q\bar{q}$ pairs.
We then compare our results to the ones obtained with a standard Glauber model initialization.
We find that regardless of the initial condition the final stage of the system contains an abundant percentage of
$q\bar{q}$ pairs; moreover spanning the possible coupling from weak to strong we find that
unless the coupling is unrealistically small, both production rate and final percentage of fermions 
is quite large.
\end{abstract}

\begin{keyword}
Quark-gluon plasma
\sep Relativistic transport theory.

\PACS 12.38.Mh\sep 25.75.Nq

%% MSC codes here, in the form: \MSC code \sep code
%% or \MSC[2008] code \sep code (2000 is the default)

\end{keyword}

\end{frontmatter}

%% \linenumbers

%% main text
\section{Introduction}
In the last decade it has been reached a general consensus that Ultra-relativistic heavy-ion collisions (uRHICs) 
at the Relativistic Heavy-Ion Collider (RHIC) and the Large Hadron Collider (LHC) create a hot and dense 
strongly interacting quark and gluon plasma (QGP)   \cite{STAR_PHENIX, ALICE_2011, Science_Muller,Fries:2008hs}. 
A main discovery has been that the QGP has a very small shear viscosity to density entropy, $\eta/s$,
which is more than one order of magnitude smaller than the one of water~\cite{Csernai:2006zz,Lacey:2006bc}, 
and close to the lower bound of $1/4\pi$ conjectured for systems at infinite strong coupling \cite{Kovtun:2004de}. 
According to the standard picture of ultrarelativistic heavy ion collisions, before the collision the two colliding
nuclei can be represented as two thin sheets of color-glass condensate~\cite{McLerran:1993ni,McLerran:1993ka,McLerran:1994vd}
which produce, immediately after the collision, a configuration of strong longitudinal chromoelectric and chromomagnetic fields 
named the glasma, see~\cite{Gelis:2010nm,McLerran:2008es,Gelis:2012ri} for reviews.

An interesting problem of uRHICs is the dynamical evolution 
of the high energy system made mainly of gluons, obtained from the decay of the glasma flux tubes,
to a locally thermalized and eventually chemically equilibrated quark-gluon plasma.
This problem has been discussed previously in \cite{Gelis:2005pb} where
quark-antiquark production rate is computed by means of the solution of the Dirac equation
in the background of the strong initial glasma field, and in \cite{Scardina:2013nua}
by simulations based on relativistic transport theory (RTT) which is a fruitful theoretical tool to study the evolution
of QGP produced in heavy ion collisions
\cite{Ruggieri:2013ova,Xu:2008av,Xu:2007jv,Bratkovskaya:2011wp,
Ferini:2008he,Plumari_BARI,Ruggieri:2013bda,Scardina:2013nua,Plumari:2012xz,Plumari:2012ep}.
In this article we follow closely Ref.~\cite{Scardina:2013nua}
and in some sense the present study is a continuation of~\cite{Scardina:2013nua}: in fact 
in~\cite{Scardina:2013nua} several aspects have not been investigated, in particular the
role of the initial non-equilibrium distribution on $q\bar q$ production times and on the chemical equilibration
of the QGP, as well as a study of the coupling dependence of the aforementioned quantities. These are the aspects
we study in the present article.

We are interested to compute the $q\bar q$ production rate initializing simulations
by a pure gluon plasma with the spectrum computed within the classical Yang-Mills (CYM) theory of the glasma.
The evolution of the initial condition is then achieved by relativistic transport theory.
The simulations with the CYM spectrum are started at $\tau_0=0.2$ fm/c, hence assuming 
in this time range the initial longitudinal gluon fields
have decayed (the decay time is of the order of $1/Q_s$ with $Q_s$ corresponding to the saturation scale) and to
populate the transverse momentum space (in the initial glasma the fields are purely longitudinal hence
the $p_T-$spectrum at $\tau=0^+$ is zero). Within our approach we cannot discuss if and how quarks are
produced before $\tau=\tau_0$; however we find that even neglecting the possible quark formation for $\tau<\tau_0$
the QCD inelastic processes are efficient enough to obtain a final state which consists mainly of $q\bar q$ pairs rather than gluons.

We also consider the problem of the chemical equilibration of the QGP. 
For a system of quarks and gluons thermalized at the same temperature $T$ the equilibrium value for
$R\equiv (N_q + N_{\bar q})/N_g$ is given by
\begin{equation}
R_{eq}=\frac{9}{4}\frac{m_q^2(T) K_2(m_q/T)}{m_g^2(T)K_2(m_g/T)}~;
\label{eq:opq}
\end{equation}
for example if quarks and gluons are massless 
one finds easily $R_{eq}=9/4$ at chemical equilibrium. 
In the case of finite masses the ratio $R$ depends on temperature even if quasiparticle
masses are temperature independent.
We find that changing the initial distribution from a Glauber model
to a CYM spectrum has some effect on the chemical equilibration, the effect being more relevant in the case
of weak coupling.

In the present study we use the quasiparticles to identify quarks and gluons propagating in the QGP.
In particular we consider the case of massless quasiparticles, which corresponds to a fluid with the perfect
massless gas equation of state; the case of massive quasiparticles with temperature independent masses,
corresponding to a fluid with the equation of state of an ideal massive gas. Finally we also consider
the case of energy density dependent quasiparticle masses, with masses 
fixed by requiring the equation of state of the fluid is the one of QCD as it is measured on the 
lattice~\cite{Plumari:2011mk}. 
We consider three quark flavors in our simulations. 
Finally, we make use of RTT with a Boltzmann kernel type with $2\rightarrow2$ processes,
see Eq.~\eqref{eq:CI1},
hence neglecting the quantum nature of particles
as well as bremsstrahlung processes.
A comment on the latter follows: 
in the bottom-up scenario \cite{Baier:2000sb}
one would expect a significant increase of soft gluons due to gluon radiation; 
however it has been shown in \cite{El:2007vg} that at RHIC and LHC conditions  
the direct and inverse processes nearly balance in the bulk,
and the yield of gluons is not very affected by bremsstrahlung. 
Of course a quantitative study has to be done in order to give a firm statement about this important point,
hence we will address the problem of soft gluon production by bremsstrahlung in the near future.

The plan of the article is as follows. In Section II we review relativistic transport theory which
is the base of our simulation code. In Section III we summarize our results concerning formation time
of quarks as well as chemical equilibration of the QGP, when we initialize simulations starting either from
a Glauber model or a CYM spectrum; in this Section we consider the cases of massless as well as massive 
quasiparticles, in the latter case considering both constant and temperature dependent masses. Finally in
Section IV we draw our conclusions.

\section{Relativistic Transport Theory}
%\subsection{Relativistic Transport Theory}
In our study we simulate the evolution of the QGP fireball by means of a code based on the 
solution of the relativistic kinetic Boltzmann-Vlasov equation for the gluon and quark distribution
function,
\begin{equation}
\left[
p^\mu \partial_\mu + M(x)\partial_\mu M(x)\nabla^\mu 
\right]f(x,p)=C[f]~,
\label{eq:BV1}
\end{equation} 
where $\nabla^\mu = \partial^\mu_p$ denotes the gradient with respect to momenta and $M(x)$ corresponds to
the quasiparticle mass: we have made the dependence of this mass on the space-time coordinates explicit,
which comes from the fact that in quasiparticle models $M$ is a function of energy density~\cite{Plumari:2011mk}
and the latter changes in space and time due to the expansion as well as to the interactions. 
It is useful to remind at this point that the use of 
a quasiparticle mass in the above equation implies that the QGP fluid has a given equation of state: 
if we take $M=0$ it means we simulate a fluid with
a perfect gas equation of state $\varepsilon = 3P$; on the other hand if $M$ is fixed with a temperature dependence according to~\cite{Plumari:2011mk}
then it is equivalent to state that the QGP fluid evolves with the lattice QCD equation of state
once the local kinetic equilibrium is reached.

In Eq.~\eqref{eq:BV1} the collision integral takes the form
\begin{equation}
C[f]=\int_2 \int_{2^\prime} \int_{1^\prime}
\left(
f_{1^\prime}f_{2^\prime} - f f_{2}
\right)
|{\cal M}|^2
\delta^4 (p + p_2 - p_{1^\prime} - p_{2^\prime})
d\Gamma_2 d\Gamma_{1^\prime} d\Gamma_{2^\prime}~,
\label{eq:CI1}
\end{equation}
where ${\cal M}$ denotes the invariant amplitude for the process $12\rightarrow 1^\prime 2^\prime$ and 
$2E_i d\Gamma_i = d^3\bm p_i/(2\pi)^3$. Both elastic as well as inelastic processes contribute to the
invariant amplitude. In this work we follow~\cite{Scardina:2013nua} and we consider only the two body processes.
In particular all the elastic cross sections we take have the form
\begin{equation}
\frac{d\sigma_{elastic}}{dt}\propto\frac{\alpha_s}{(t-m_D^2)^2}~,
\label{eq:gg}
\end{equation} 
where $t$ denotes the invariant Mandelstam variable and $m_D=0.7$ GeV is used 
as an infrared regulator and determines how isotropic the cross section is
(the smaller $m_D$ the more forward peaked the cross section). 
The overall factor in the previous equation changes channel by channel
($q\bar{q}$, $qq$, $qg$, $\bar{q}g$ and $gg$)
as in~\cite{Scardina:2013nua}.

In this study we fix the value of $\alpha_s$ and compute the relevant cross sections.
In order to estimate an appropriate value of $\alpha_s$ we use 
the total cross section for $gg$ scattering among massless gluons
and require that the value of $\eta/s$ obtained by this massless gluon-gluon scattering is approximately the one used in 
hydrodynamics and/or transport simulations. To estimate such a coupling we use the approximate kinetic theory 
relation~\cite{Plumari:2012xz,Plumari:2012ep}
\begin{equation}
\sigma_{gg}^{massless} = \frac{3}{10}\frac{T}{n}\frac{1}{\eta/s}~,
\end{equation}
which, for a reference temperature of $T=0.3$ GeV gives $\alpha_s = 0.7$ for
$\eta/s=2/4\pi$ as used in heavy ion collisions 
simulations~\cite{Ruggieri:2013ova,Ruggieri:2013bda,Schenke:2012wb}. 
A value of $\alpha_s \approx 0.9$ would be required for $\eta/s=1/4\pi$, which certainly
would cause the gluons-to-quarks processes to be even faster but which does not affect considerably the
final quark spectra as well as the final gluon number to quark number ratio,
see next Section. We will also consider the case
of a running strong coupling,
\begin{equation}
\alpha_s(T)=\frac{2\pi}{9}\frac{1}{\log(\pi T/\Lambda)}~,
\label{eq:poi}
\end{equation}
with $\Lambda=0.2$ GeV; for example at $T=0.3$ GeV one finds $\alpha_s\approx 0.4$.

For the inelastic processes between quarks and gluons $g g \leftrightarrow q\bar{q}$ we have
${\cal M} = {\cal M}_s + {\cal M}_t + {\cal M}_u$. For the massless case the cross sections
for such processes are the textbook pQCD cross section for jet
production in high-energy proton-proton collisions. With massive
quarks the calculations are the Combridge cross sections~\cite{Combridge:1978kx,Biro:1990vj} 
used to evaluate heavy-quark production. In our case we
have considered a finite mass for both gluons and quarks together
with a dressed gluon propagator. 
It is worth to mention here that the cross section for the process at hand 
is dominated by the $t$-
and $u$-channel and their interference, while the $s$-channel alone is
negligible. The squared matrix element of the $t$-channel is given
by
\begin{equation}
|(t-m_q^2){\cal M}_t|^2 = \frac{8}{3}\pi^2\alpha_s^2
\left[
(m_q^2 - t)(m_q^2 - u) -2m_q^2(t + m_q^2) -4 m_q^2 m_g^2 -m_g^4
\right]~,
\label{eq:2}
\end{equation}
where $m_q$, $m_g$ denote the quasiparticle masses for quarks and gluons respectively.

\section{Results}

\subsection{The massless case}

In the present work we focus on two different initial conditions for our simulations.
The first one is named thermal
Glauber (Th-Glauber) in which we distribute particles in transverse coordinate space according to
the usual Glauber model with the standard mixture $0.85 N_{part} + 0.15 N_{coll}$, while distributing
particles in boost invatiant way along the longitudinal direction;
initial momentum space distribution corresponds to a thermal distribution in transverse momentum space
and a boost invariant distribution in the longitudinal direction with $y=\eta$, with $y$ and $\eta$
denoting momentum and space-time rapidities respectively.
The other initial condition we use in our simulations is named CYM initialization, in which 
transverse plane particle distribution corresponds to a Glauber model with $N_{coll}$
scaling~\cite{Schenke:2012wb} while initial $p_T-$distribution is obtained by the solution
of classical Yang-Mills equations~\cite{Schenke:2013dpa}; also in this case we assume longitudinal
boost invariance and $y=\eta$.

\begin{figure}[t!]
\begin{center}
\includegraphics[width=9cm]{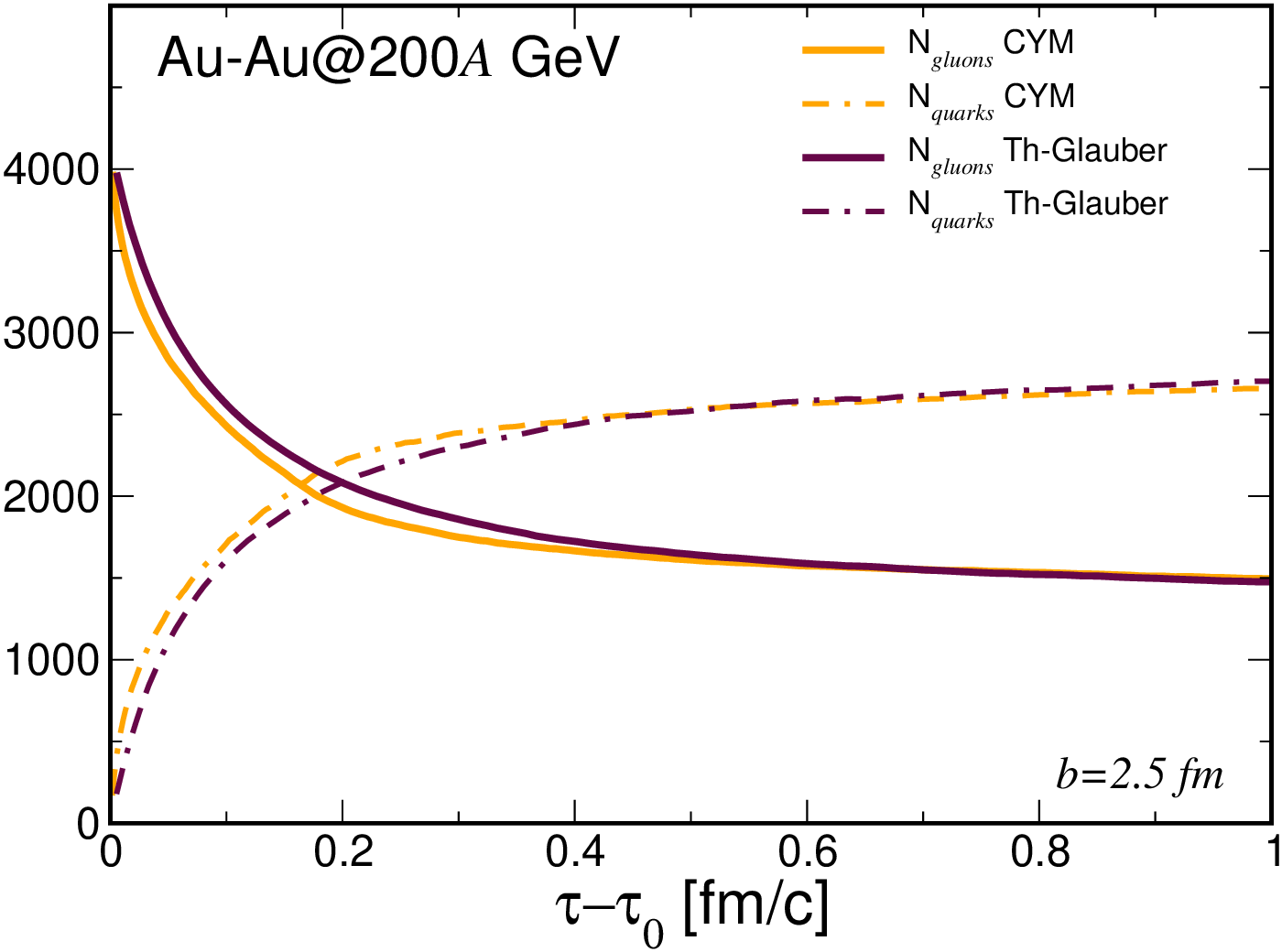}
\caption{\label{Fig:1}Early time evolution of particle number for thermal Glauber (indigo)
and CYM (orange) initializations. Initial gluon number corresponds to $dN_g/dy=$1040.}
\end{center}
\end{figure}

In Fig.~\ref{Fig:1} we plot the time evolution of the particle number for the several initializations
we consider in this work. 
On the horizontal axis we measure the time difference $\tau-\tau_0$
where $\tau_0$ corresponds to the initialization time;
in this study we chose $\tau_0 = 0.6$ fm/c for the case of the Glauber initialization and $\tau_0=0.2$ fm/c
for the CYM initialization. In the latter case we assume a smaller initialization time because
we do not need to assume any pre-thermalization, which instead is assumed in the Glauber initialization.
In the figure, $N_{quarks}$ corresponds to the total number of quarks {\em plus} the total number of antiquarks.
The results we present in this article correspond to simulations run for Au-Au collisions at RHIC
energy, for $b=2.5$ fm; the initial gluon multiplicity is $dN_g/dy=$1040.

\begin{figure}[t!]
\begin{center}
\includegraphics[width=9cm]{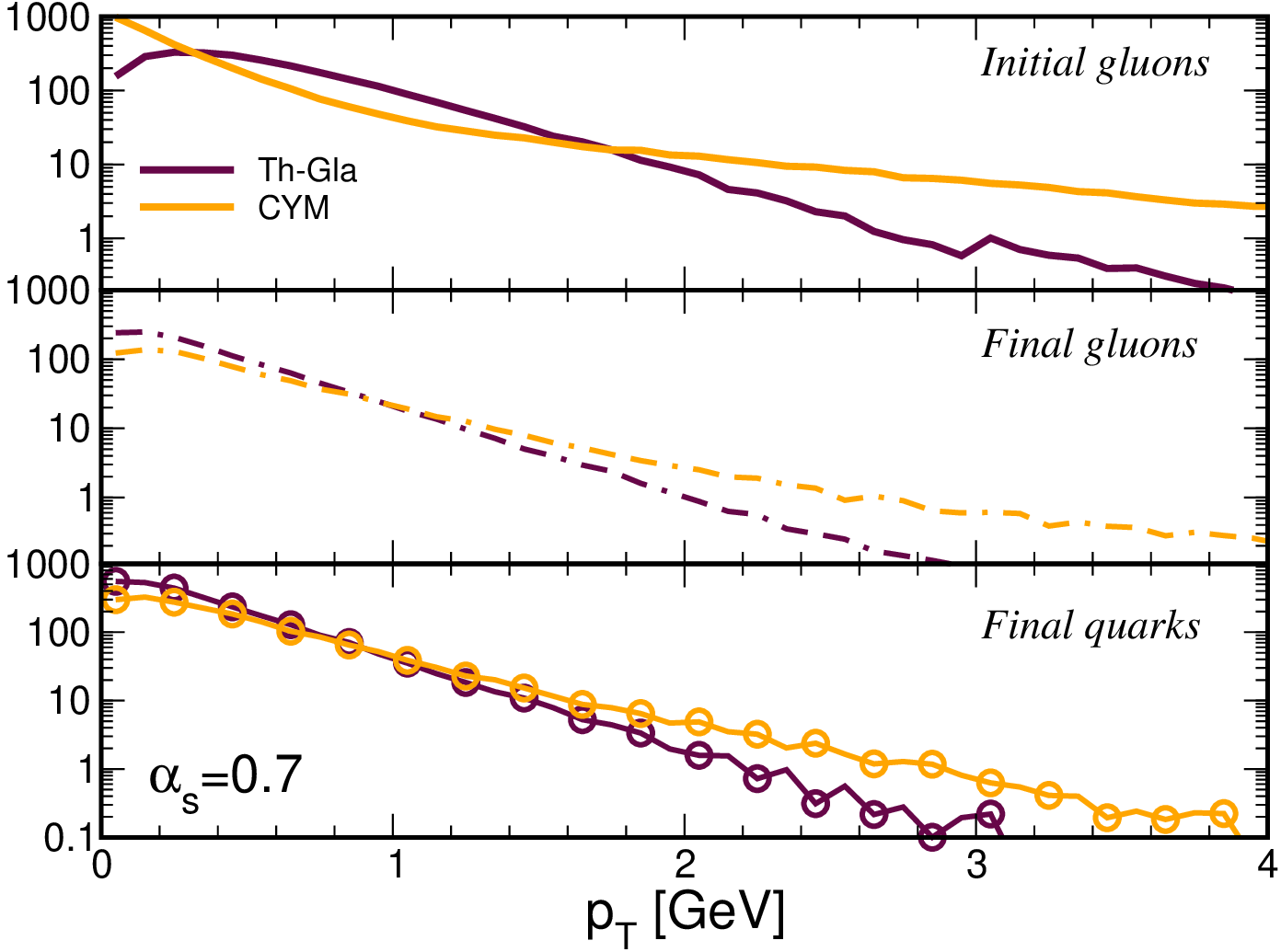}
\caption{\label{Fig:2}Time evolution of particle spectra for thermal Glauber (indigo)
and CYM (orange) initializations. In both simulations the initial gluon number
corresponds to $dN_g/dy=$1040.}
\end{center}
\end{figure}

In Fig.~\ref{Fig:2} we plot the results for initial and final particle spectra at midrapidity $|y|\leq 0.5$. 
In the figure we do not show initial spectra for quarks because we assume that at $\tau=\tau_0$ there are no quarks
in the system. This might be a crude approximation for the Glauber initialization since it is equivalent to assume
that some dynamics took place up to $\tau=\tau_0$ to thermalize the system in the transverse plane,
nevertheless affecting only gluons and not producing any quark; nevertheless for the early times initializations
this might be not a so bad approximation since we are only assuming that within a time of the order of $1/Q_s\approx\tau_0$
the system is mainly made of gluons, then as soon as we initialize the dynamics the QCD processes cause the conversion
of gluons to quarks. We remark that a more realistic condition including a finite number of quarks
can only make stronger our final result that quark production is quite fast and abundant. 

The results summarized in Figg.~\ref{Fig:1} and~\ref{Fig:2} are obtained assuming a value of $\alpha_s = 0.7$.
As a reference we remind that 
this value of the coupling corresponds to a value of $\eta/s\approx 2/4\pi$ at temperature $T=0.3$ GeV,
which is not far from the value of $\eta/s$ which is normally used in ultrarelativistic heavy ion collisions simulations
to quantify the dissipation of the QGP fluid.

\begin{figure}[t!]
\begin{center}
\includegraphics[width=7cm]{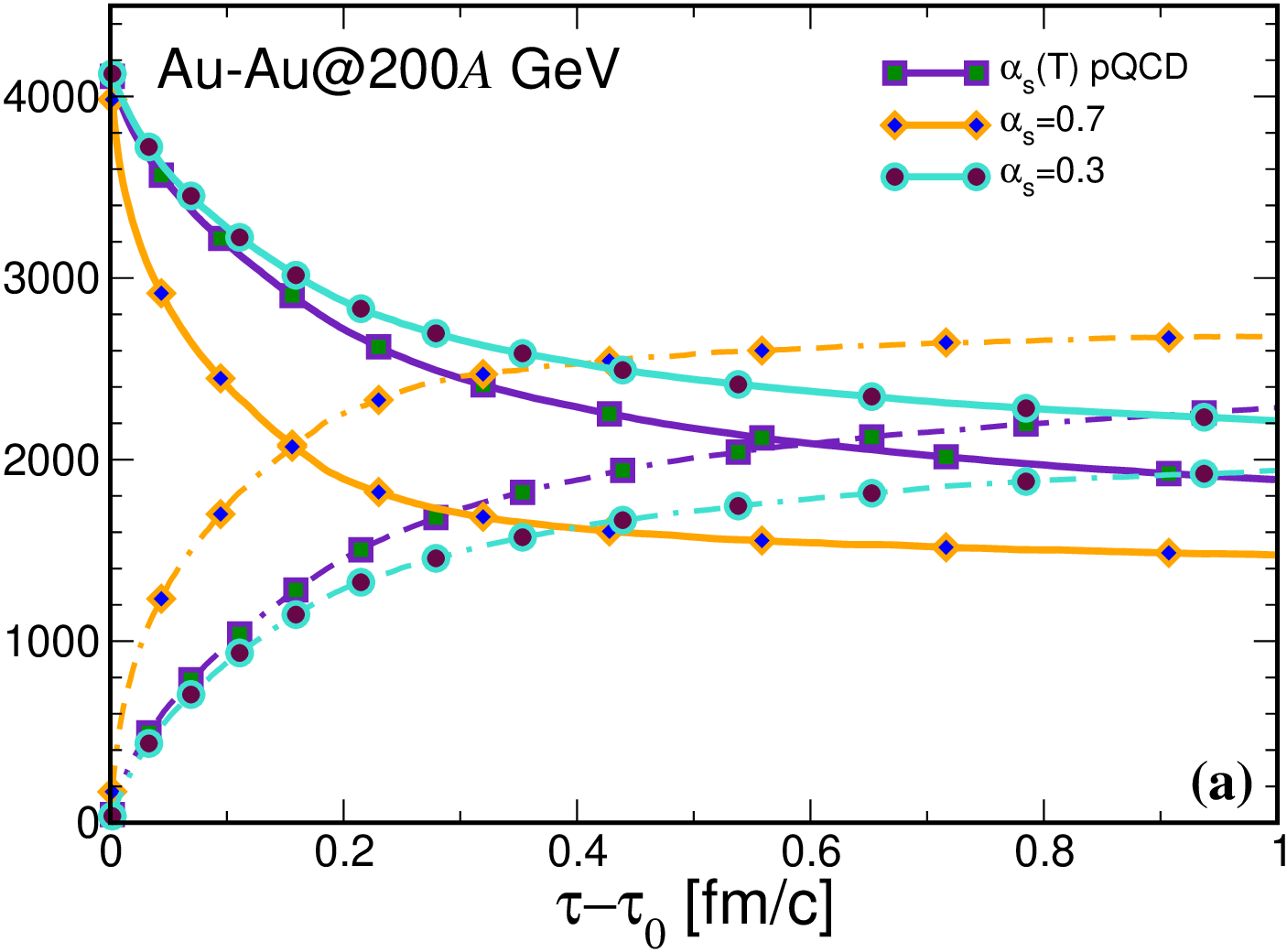}~~~
\includegraphics[width=7cm]{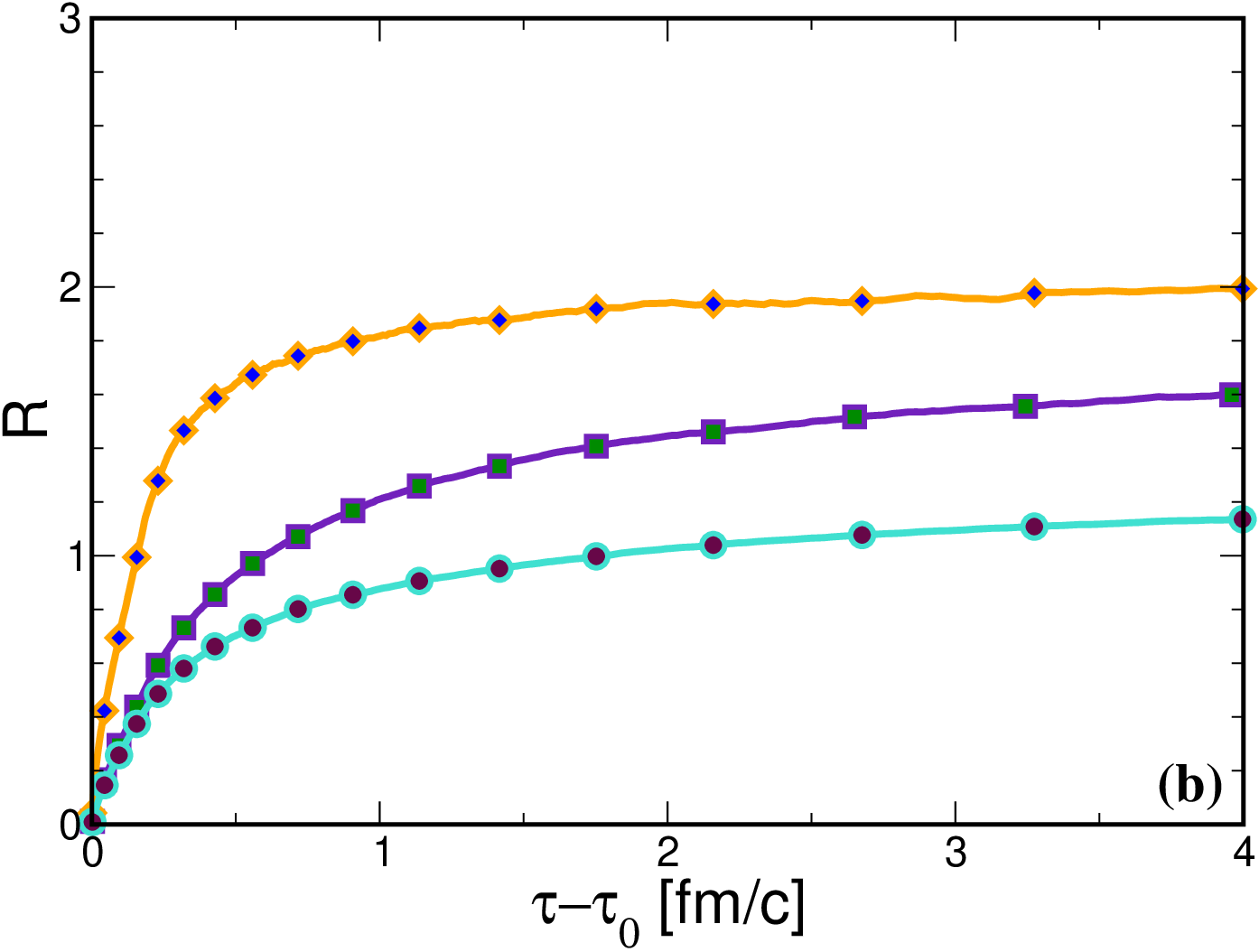}
\caption{\label{Fig:3}{\em Left panel}: Early time evolution of gluons and fermions number for the case
of CYM initialization and several choices for the coupling. {\em Right panel}: ratio of total quark number over
gluon number as a function of time. Data style convention in the right panel is the same as the one in the
left panel. Calculations correspond to Au-Au collision at $\sqrt{s}=200A$ GeV and $b=2.5$ fm.}
\end{center}
\end{figure}

In the left panel of Fig.~\ref{Fig:3} we plot the gluon and total quark numbers as a function of time, 
focusing on the CYM initialization and on the early stage of the time evolution of the fireball.
For other initializations we obtain similar results which are not shown for space sake.
In the figure we have shown three different cases corresponding to three different choices for $\alpha_s$.
In particular, diamonds and circles correspond to a fixed value of $\alpha_s$, namely $\alpha_s=0.7$
and $\alpha_s=0.3$ respectively; on the other hand squares correspond to the case of the temperature
dependent $\alpha_s$
given by Eq.~\eqref{eq:poi}. 

The definition of a quark formation time $\tau_q$ is quite arbitrary, therefore 
we define it as the time necessary to have a mixture made of 
$50\%$ of quarks plus antiquarks and $50\%$ of gluons, corresponding to $R=1$
with $R=(N_{q}+N_{\bar q})/N_g$. 
As already discussed in the case $\alpha_s=0.7$ 
the system is very efficient in converting gluons to quarks,
in fact the total number of quarks becomes comparable to the one of gluons within $\tau_q\approx0.15$ fm/c.
It is interesting that in the case of the perturbative running coupling, indigo data in Fig.~\ref{Fig:3},
even if the average coupling is smaller than the case $\alpha_s=0.7$ (of about a factor of 1.5 when averaged
on the fireball) the system is still quite efficient in producing quarks via the QCD inelastic processes.
In fact in this case we find $\tau_q\approx0.6$ fm/c, which is about a factor of 4 larger than the time required
in the case  $\alpha_s=0.7$ but still very fast compared to the time evolution of the fireball. Finally in the case
$\alpha_s=0.3$  we find that quark production is quite slow compared to the previous cases, having 
$\tau_{q} \approx 2$ fm/c which is approximately 3.3 times larger than $\tau_q$ 
obtained within the perturbative running coupling and approximately $1/2$ of the lifetime of the QGP in the fireball.

In the right panel of Fig.~\ref{Fig:3} we plot the ratio $R=(N_{q}+N_{\bar q})/N_g$ again for the case of the
CYM initialization, for the same values of $\alpha_s$ considered in the left panel of the same figure. 
In the case of $\alpha_s=0.7$ the ratio $R$ needs a short time range of about 1 fm/c
to reach a value close to the equilibrium value for a massles gas, $R=9/4$. 
On the other hand in the case of a running perturbative coupling,
even if the quark production has been efficient in the early stages after the collision, the system is not able to
chemically equilibrate onto the chemical equilibrium value within the lifetime of the QGP in the fireball; 
in fact $R$ saturates to a somewhat smaller value, $R\approx 1.7$ while $R=9/4$ at equilibrium. 
Finally in the case of weak coupling
$\alpha_s=0.3$ we find the value $R\approx 1.2$ is reached within the lifetime of the fireball,
meaning the system is unable to reach chemical equilibration in weak coupling regime even though quark
fraction is still abundant. The different asymptotic behaviours in this case of massles
quasiparticles are due to the freezout which freezes the interactions hence arresting conversion
processes from quarks to gluons.

\subsection{The massive case: temperature independent masses}

\begin{figure}[t!]
\begin{center}
\includegraphics[width=7cm]{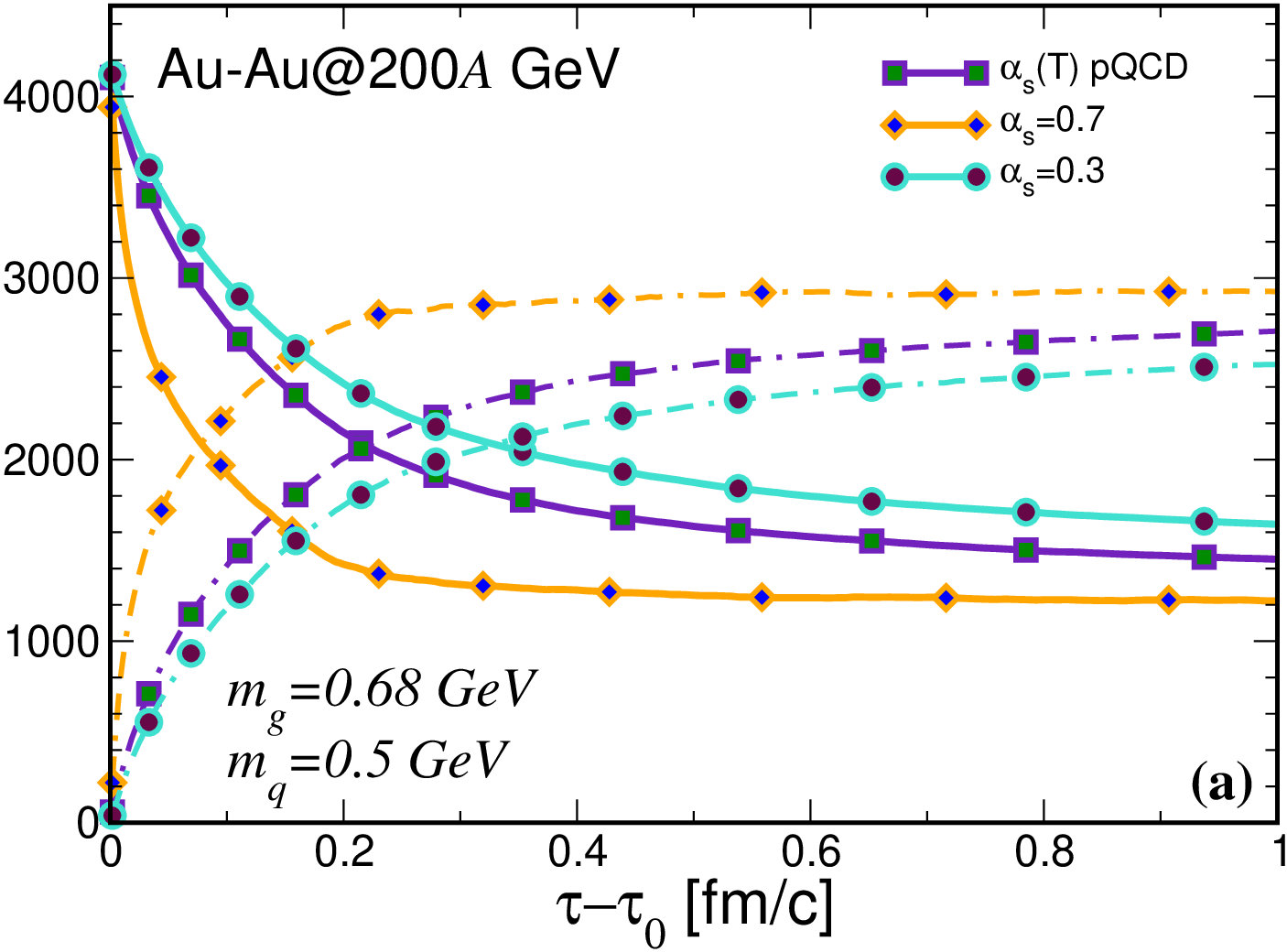}~~~
\includegraphics[width=7cm]{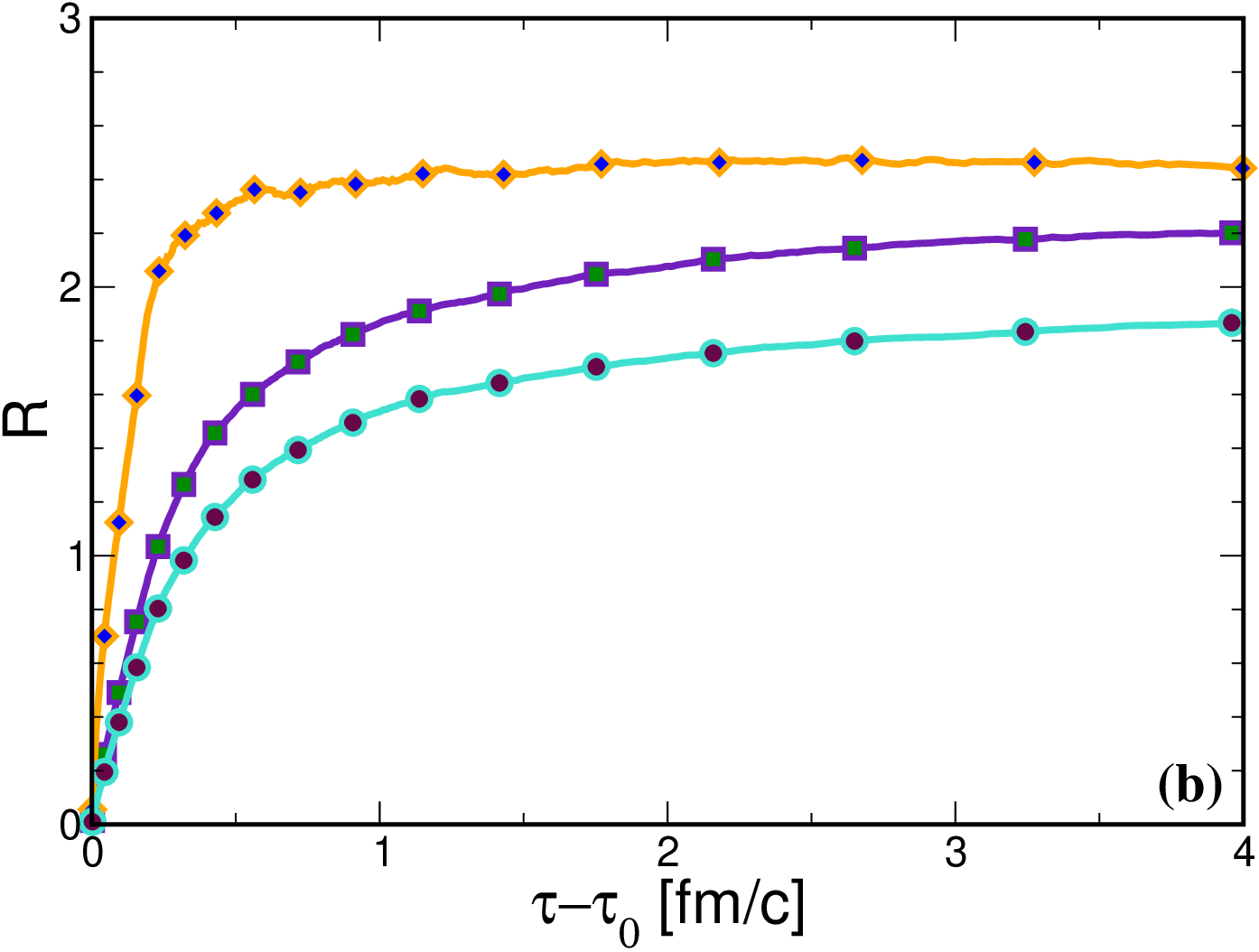}
\caption{\label{Fig:4}{\em Left panel}: Early time evolution of gluons and fermions number for the case
of CYM initialization and several choices for the coupling, for the 
case of massive quasiparticles. {\em Right panel}: ratio of total quark number over
gluon number as a function of time. Data style convention in the right panel is the same as the one in the
left panel. Calculations correspond to Au-Au collision at $\sqrt{s}=200A$ GeV and $b=2.5$ fm.}
\end{center}
\end{figure}

In the left panel of Fig.~\ref{Fig:4} we plot the gluon and total quark numbers as a function of time,
for the case of massive quasiparticles. In order to emphasize the role of a finite mass for the quasiparticles,
as well as of the different mass among quarks and gluons, we have restricted ourselves to the case of temperature
independent masses. According to the results of~\cite{Plumari:2011mk} we have chosen the ratio of the gluon to quark 
mass to be $3/2$, neglecting a mild temperature dependence of the masses in the range of temperature
relevant for the evolution of the fireball at RHIC energy. 
We have chosen the quark mass to be $m_q=0.5$ GeV in agreement with the results of~\cite{Plumari:2011mk}.
We find that regardless of the choice for the coupling, considering massive quasiparticles amounts to
increase the quarks to gluons number ratio $R$ of about $30\%$
along with a general acceleration of $\tau_q$, compare Figg.~\ref{Fig:3} and~\ref{Fig:4}. 
This is easily understood because
in this case quarks are lighter than gluons and statistically it is easier to convert a pair of gluons 
into a $q\bar{q}$ pair.

\subsection{The massive case: temperature dependent masses}

\begin{figure}[t!]
\begin{center}
\includegraphics[width=7cm]{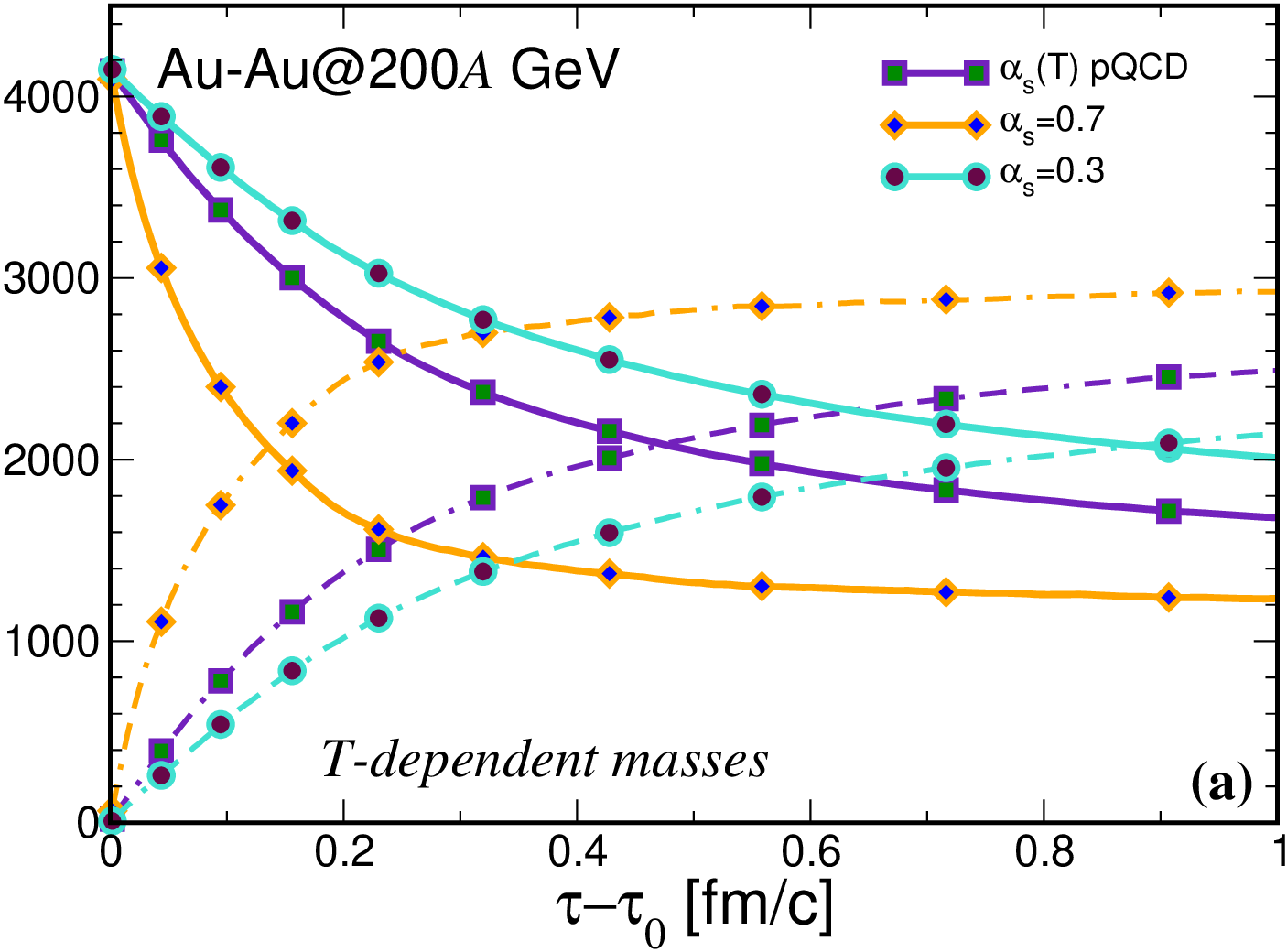}~~~
\includegraphics[width=7cm]{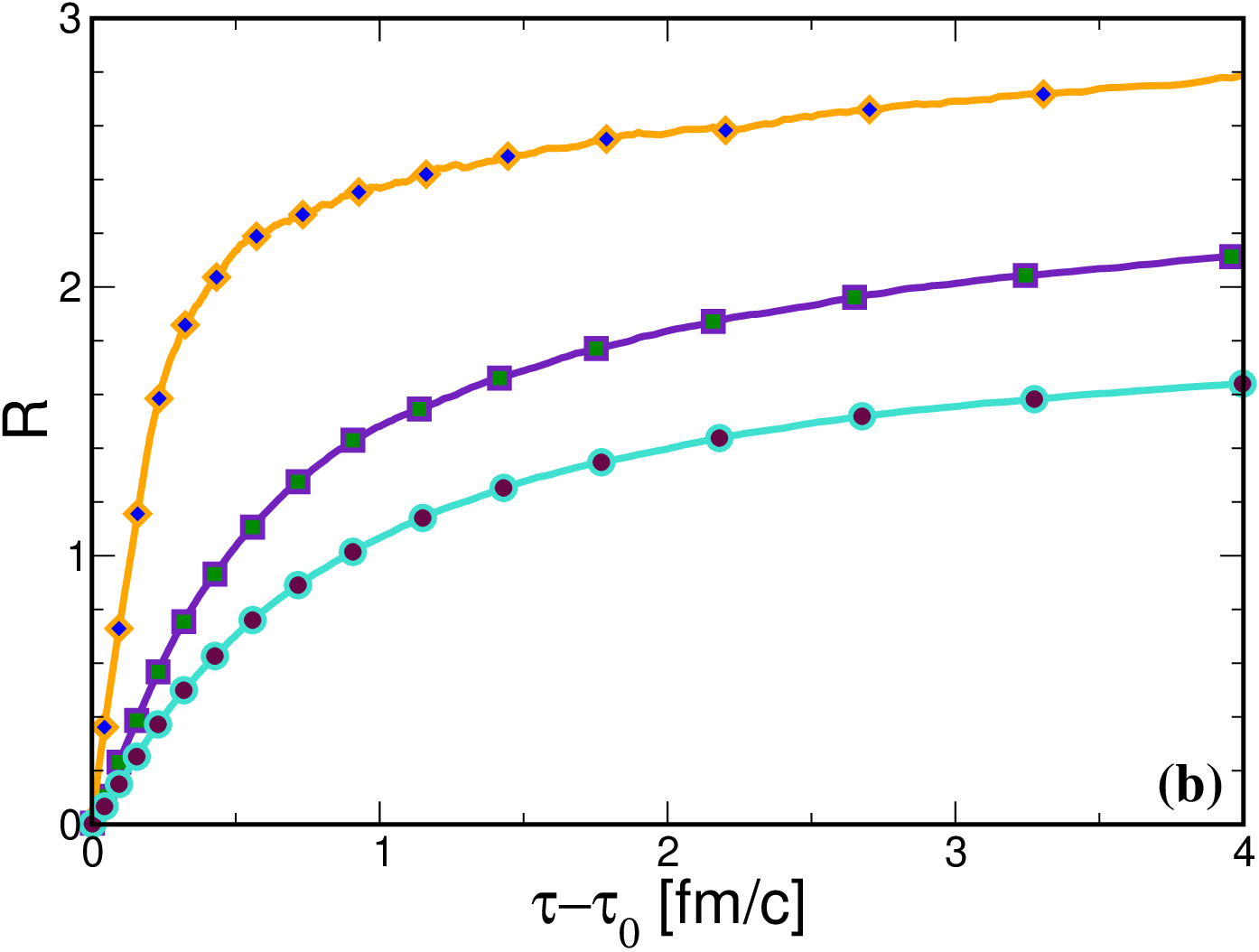}
\caption{\label{Fig:5}{\em Left panel}: Early time evolution of gluons and fermions number for the case
of CYM initialization and several choices for the coupling,
for the case of $T-$dependent masses. {\em Right panel}: ratio of total quark number over
gluon number as a function of time. Data style convention in the right panel is the same as the one in the
left panel. Calculations correspond to Au-Au collision at $\sqrt{s}=200A$ GeV and $b=2.5$ fm.}
\end{center}
\end{figure}

In the left panel of Fig.~\ref{Fig:5} we plot the gluon and total quark numbers as a function of time,
for the case of massive quasiparticles with a temperature dependent mass. 
The quasiparticle masses are computed according to~\cite{Plumari:2011mk} in order
to fit the lattice thermodynamics data; in other words, the masses we consider in this section
are such that the equation of state of the fluid we simulate is that of lattice QCD.
Left panel of Fig.~\ref{Fig:5} focuses on the very early history of the fireball; 
we find that this case does not differ in a relevant way from the case of temperature
independent masses discussed in the previous subsection. In fact also in this case
perturbative QCD processes lead to fast quarks production. 
However a considerable quantitative difference is found for the final values of the
ratio $R$ of quarks to gluons, which is more evident in the case of strong coupling
$\alpha_s=0.7$. Indeed in the case of temperature independent masses of Fig.~\ref{Fig:4}
we found that $R$ approaches $\approx2.4$ within few fm/c; instead in the case
of temperature dependent masses we find that $R\approx 3$ at the end of the evolution.

On the other hand, for the running coupling constant case, circles in  Fig.~\ref{Fig:5},
we find that the use of the temperature dependent masses does not affect very much
the result obtained within fixed masses. This can be understood {\em a posteriori} since
the quark and gluon masses in Fig.~\ref{Fig:4} do not differ too much from those of Fig.~\ref{Fig:5},
besides the peripheral cells where the temperature is smaller thus the masses are larger,
but the coupling is smaller than the case $\alpha_s = 0.7$ (roughly corresponding to
$\eta/s = 0.15$) implying the system 
is less effective in converting gluons to quarks. The regions of the fireball where temperature
is close to the critical temperature are still important, because the smaller the temperature
the larger the ratio $R$ in this case,   but these regions can be effective in the production
of quarks only if the coupling is large enough, which is achieved by the case $\alpha_s = 0.7$
but non in the case of the running perturbative coupling. This explains the results on the right
panel of Fig.~\ref{Fig:5}.

\subsection{CYM versus Glauber initializations}  

\begin{figure}[t!]
\begin{center}
\includegraphics[width=8cm]{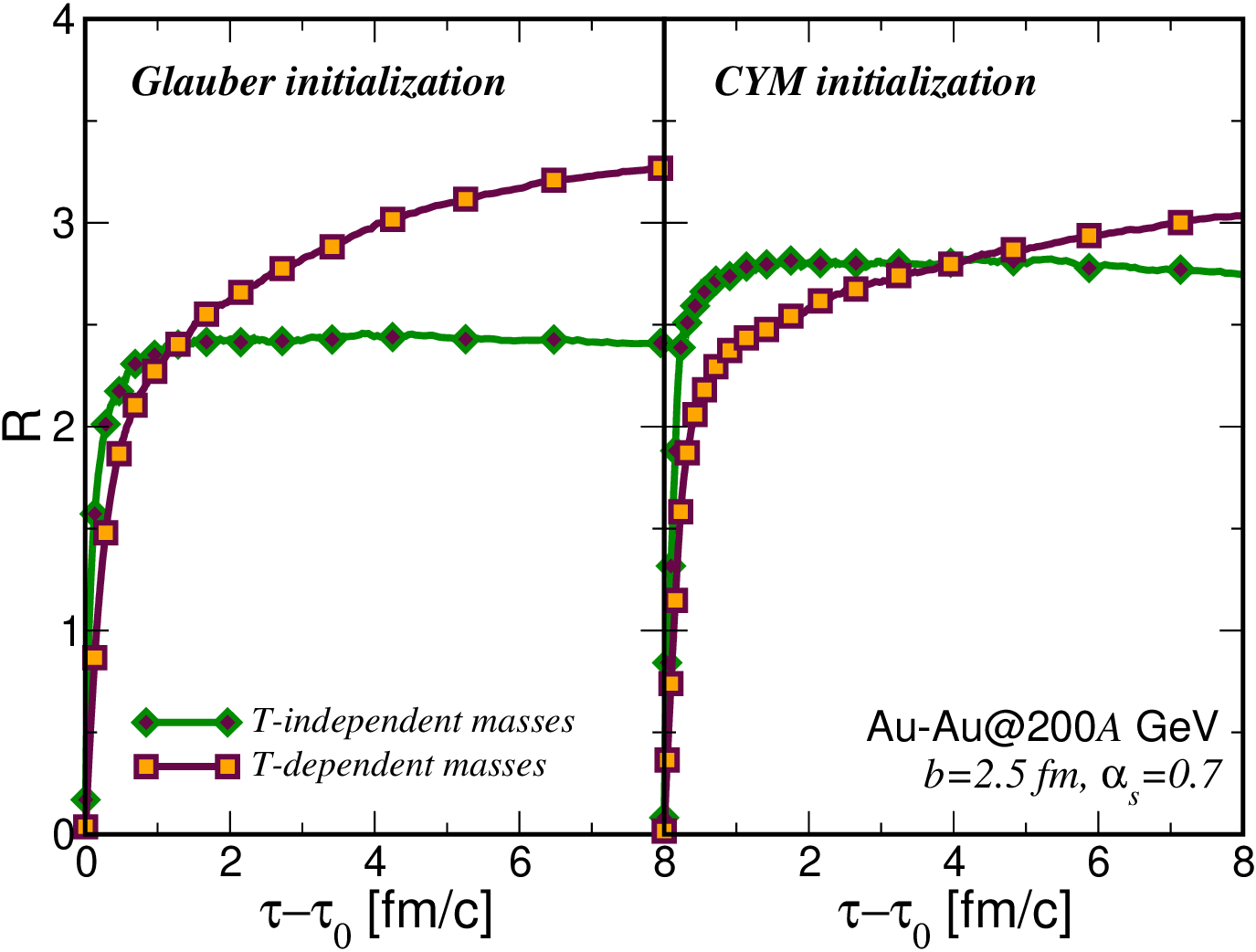}
\caption{\label{Fig:6}Comparison of the ratio $R$ for the Glauber and the CYM initializations.
Left panel corresponds to Glauber initialization, right panel to CYM initialzation.
Calculations are performed with a fixed $\alpha_s=0.7$,
for Au-Au collision at $\sqrt{s}=200A$ GeV and $b=2.5$ fm.}
\end{center}
\end{figure}
 
\begin{figure}[t!]
\begin{center}
\includegraphics[width=8cm]{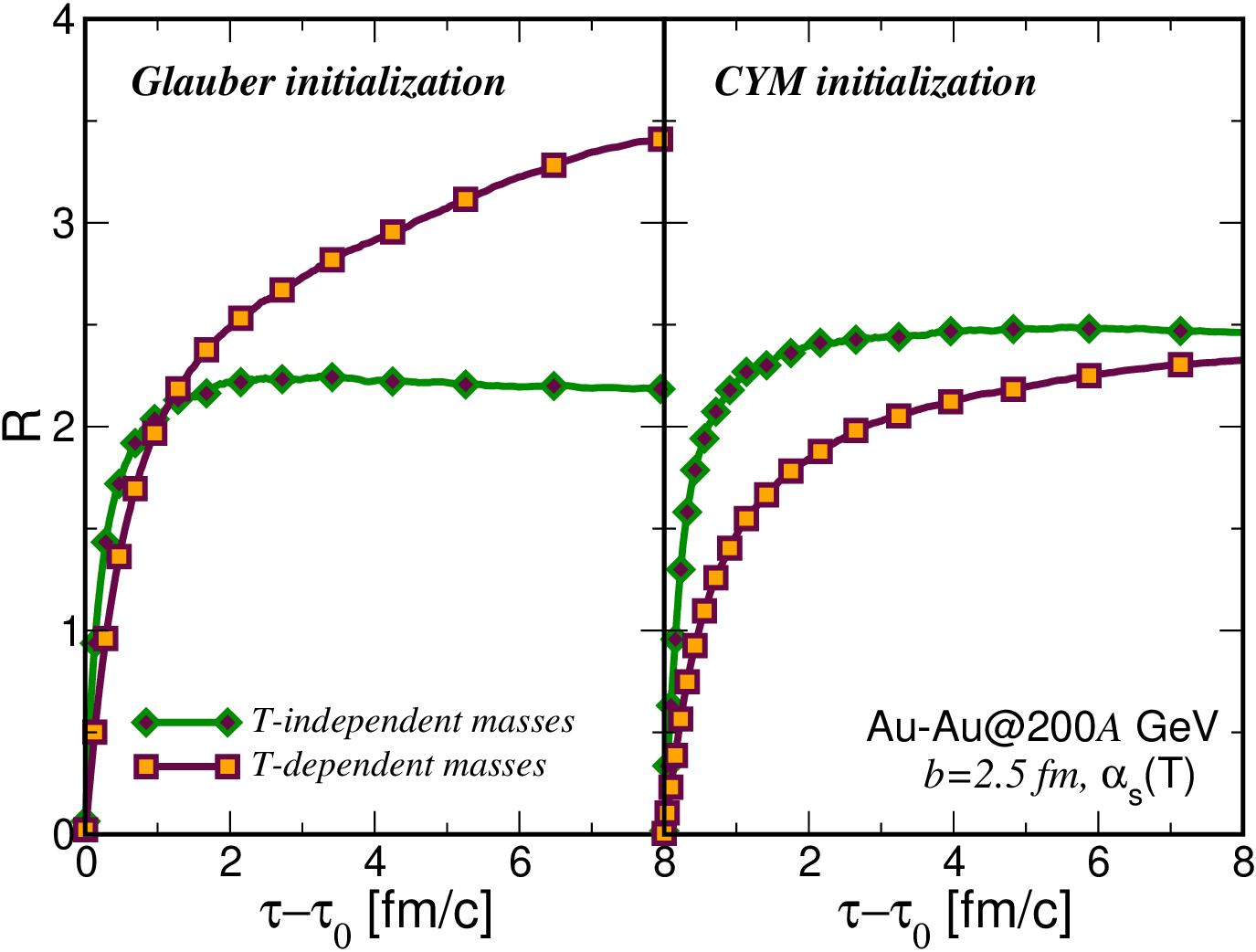}
\caption{\label{Fig:6b}Comparison of the ratio $R$ for the Glauber and the CYM initializations.
Left panel corresponds to Glauber initialization, right panel to CYM initialzation.
Calculations are performed with a temperature dependent $\alpha_s$,
for Au-Au collision at $\sqrt{s}=200A$ GeV and $b=2.5$ fm.}
\end{center}
\end{figure}

In this subsection we compare the evolution of the quarks to gluons ratio obtained by the Glauber
and the CYM initializations, for both cases of strong fixed and temperature dependent couplings.
The cases of both temperature dependent and constant masses are considered. This comparison is instructive
since it permits to enlighten about the role of the different initialization on quark production.
It is useful to remind here that the case of temperature dependent masses corresponds to assume
the QGP fluid is characterized by the equation of state of lattice QCD.

In Fig.~\ref{Fig:6} we plot the evolution of $R$ as a function of time for the Glauber initialization
(left panel) and the CYM initialization (right panel). The results are shown for the cases of constant mass and
temperature dependent mass, with perturbative running and fixed strong coupling (the latter corresponding
to $\alpha_s=0.7$). In the case of strong coupling, left panel of  Fig.~\ref{Fig:6}, we find that for the Glauber initialization switching from a 
constant mass to a thermal one leads to an increase of the ratio $R$ at later times. This is mainly due to the fact
that the number of gluons decreases  (and that of quarks increases) continuously in the case of a temperature
dependent mass, while it saturates after a time of about $2$ fm/c in the case of a constant mass.
We find the same result for the CYM initialization, even if in this case the effect of turning on a 
temperature dependent mass is milder. However, the conclusion is that regardless the initial condition chosen
in the simulation, at later times the fireball consists of about $80\%$ of quarks and only $20\%$ of gluons
even starting from a condition in which the system is made of gluons only. 
Finally in Fig.~\ref{Fig:6b} we plot the ratio $R$ for the case of $\alpha_s(T)$.

\section{Conclusions}

In this Article we have studied quark production from a pure gluon plasma produced in ultrarelativistic heavy ion collisions.
Our main goal has been to study how a pure gluon plasma evolves dynamically to a quark-gluon plasma
by means of QCD inelastic processes;
the impact of the initial condition on the relative composition in quarks and gluons
and on quark production times has been investigated. 
We have considered two different kinds of initializations: the usual Glauber initialization and one based
on the gluon spectrum obtained by the classical Yang-Mills theory (CYM) of the glasma. 
Given the initial condition, we have evolved it according to relativistic transport theory, and quark production
has been taken into account by means of QCD inelastic processes.

The definition of a quark formation time $\tau_q$ is quite arbitrary, therefore 
we have defined it as the time necessary to have a mixture made of 
$50\%$ of quarks plus antiquarks and $50\%$ of gluons. 
We have found that the initial condition does not have a relevant impact on $\tau_q$;
this is true both if we consider a pure perturbative
coupling and a strong coupling regime in which $\alpha_s$ is chosen scale-independent and of magnitude
such that the corresponding $\eta/s$ is in the range $1\div 2$.
The coupling has some effect on $\tau_q$: changing the coupling from strong to weak we have obtained
$\tau_q$ increases about of a factor $4$; nevertheless even in weak coupling we find abundant fermion
production takes place within $\tau\leq 1$ fm/c.

We have then computed the ratio $R\equiv (N_q + N_{\bar q})/N_g$ and we have found it
is affected not only by the coupling strength but also on the kind of quasiparticle mass we implement in the simulation.
In the case of strong coupling we have found that regardless of the initial condition the ratio $R$
approaches about $2.5 \div 3.5$; in the case of weak coupling the effect of the initialization
is quite strong for the case of temperature dependent masses, being instead not so strong in the case
of massive quasiparticles with a constant mass. It is interesting that even in the weak coupling case
the asymptotic $R\approx 2  \div 2.5$   signaling an abundant production of quarks plus antiquarks
has been achieved during the fireball's lifetime.

Among the many interesting aspects which deserve further study we briefly mention here the possibility to
include, in future studies, the Bose-Einstein enhancing factors in the collision integral which might
lead to a transient condensate in the very early stages of the 
glasma evolution~\cite{Blaizot:2013lga,Blaizot:2011xf,Blaizot:2012qd,Liao:2014qoa,Scardina:2014gxa},
hence affect the conversion of gluon to quarks~\cite{Blaizot:2014jna}. 
Moreover it will be interesting to compare the quark production
in QGP studied in the present article with the one obtained by the shattering of the flux tubes via the Schwinger
mechanism, following the lines of~\cite{Ryblewski:2013eja}. Both these problems will be the subject of
forthcoming publications.

\emph{Acknowledgements.} We acknowledge B. Schenke and R. Venugopalan
for kindly providing us the data of the initial CYM spectra and for fruitful
correspondence. We also acknowledge A. Puglisi for enlightening discussions.
V. G., M. R. and F. S. acknowledges the ERC-STG funding under the QGPDyn
grant.

%% If you have bibdatabase file and want bibtex to generate the
%% bibitems, please use
%%
%%  \bibliographystyle{elsarticle-num} 
%%  \bibliography{<your bibdatabase>}

%% else use the following coding to input the bibitems directly in the
%% TeX file.

\end{document}